\documentclass[fleqn,10pt]{wlscirep}
 
\usepackage{graphicx}
\usepackage{bm}
\usepackage{xcolor}
\usepackage{url}

\usepackage{ulem} 

\begin{document}

\title{Detecting sequences of system states in temporal networks}
\author[1,*]{Naoki Masuda}
\author[2]{Petter Holme}
\affil[1]{Department of Engineering Mathematics,
Merchant Venturers Building, University of Bristol,
Woodland Road, Clifton, Bristol BS8 1UB, United Kingdom}
\affil[2]{Institute of Innovative Research, Tokyo Institute of Technology, Nagatsuta-cho 4259, Midori-ku, Yokohama, Kanagawa, 226-8503, Japan}
\affil[*]{Correspondence to naoki.masuda@bristol.ac.uk}

\begin{abstract}
Many time-evolving systems in nature, society and technology leave traces of the interactions within them. These interactions form temporal networks that reflect the states of the systems. In this work, we pursue a coarse-grained description of these systems by proposing a method to assign discrete states to the systems and inferring the sequence of such states from the data. Such states could, for example, correspond to a mental state (as inferred from neuroimaging data) or the operational state of an organization (as inferred by interpersonal communication). Our method combines a graph distance measure and hierarchical clustering. Using several empirical data sets of social temporal networks, we show that our method is capable of inferring the system's states such as distinct activities in a school and a weekday state as opposed to a weekend state. We expect the methods to be equally useful in other settings such as temporally varying protein interactions, ecological interspecific interactions, functional connectivity in the brain and adaptive social networks.
\end{abstract}

\maketitle

\section{Introduction\label{sec:introduction}}

Many systems composed of interacting elements can be represented as networks, and nowadays it can be quite easy to obtain large amounts of interactions from such a system. One challenge for network science~\cite{Newman2010book,Barabasi2016book} is to condense the information obtained from such streams of data into simplified, more narrative, pictures of what happens in the system. For networks that are static in time, this has been well studied as the problem of community detection in networks~\cite{Fortunato2010PhysRep}. However, many networks in the real world are changing over time. This change is interesting in its own right because it reflects the forces acting upon the system. Furthermore, if the timescale of network change is similar to that of dynamic processes relevant to the system (e.g., information spreading, gene expression, transportation), the dynamics \textit{of} the network might affect, and be affected by, the dynamics \textit{on} the network. As a consequence, traditional network theory developed for static networks may not apply. Coarse-graining the data (i.e., inferring the meso-scale structure of a network such as communities) is one such example, where straightforward generalizations from static networks do not work. This knowledge gap motivates the study of temporal (i.e., time-varying) networks, in which nodes and/or edges are time-stamped. We now know that the temporality of networks does change our understanding of networks in many cases and provides us with richer information and more efficient manipulative tools~\cite{HolmeSaramaki2012PhysRep,HolmeSaramaki2013book_Springer,Holme2015EurPhysJB,Masuda2016book,Masuda2017book}.

Let us consider human social behaviour as an example. Research has shown that human behaviour can often be accurately modelled by dynamical processes depending on discrete states. For example, email correspondence behaviour of individuals was modelled as a two-state point process in which an individual was assumed to switch between an active state and a normal state, supplemented by circadian and weekly modulations of the event rate \cite{Malmgren2008PNAS}. The event rate in the active rate was assumed to be larger than that in the normal state. A similar model accounted for historical letter correspondence behaviour of celebrity \cite{Malmgren2009Science}. These and other \cite{Karsai2012SciRep,Vajna2013NewJPhys,Raghavan2014IeeeTransComputSocSyst,Jiang2016NewJPhys} discrete-state models have been useful in explaining empirical data such as long-tailed distributions of inter-event times. The philosophy underlying these models is that a system such as an individual human can be modelled by a sequence of discrete states between which the system stochastically switches (akin to the dynamics in a hidden Markov model \cite{Rabiner1989IEEE,Bishop2006book}).

In the present study, we hypothesise that the state of a networked system as a whole, rather than behaviour of individual nodes, is encoded in the edges (i.e., pairwise interactions) of the network and can be summarized as a time series of state changes. An obvious example of this would be day-time versus night-time states of human interaction, with the former possibly having more frequent interaction than the latter. Another type of state change may be between weekdays and weekends \cite{Malmgren2008PNAS}. Biological, ecological, engineered and other temporal networks \cite{HolmeSaramaki2012PhysRep,Holme2015EurPhysJB} may also be characterised by switching behaviour. We show in the present study that we can distinguish discrete states of temporal networks that have similar event rates nevertheless different in internal structure of the time-stamped network. We also show that such different states are often interpretable. We refer to the inferred transitory dynamics between discrete states as \textit{system state dynamics} of temporal networks (or simply \textit{state dynamics}). To determine the state dynamics, we use distance measures for static graphs to categorise static networks at different times, which we call the snapshots, into groups; we regard each group as a state of the temporal network.

Change-point and anomaly detection for temporal networks is also concerned with detecting changes of networks over time \cite{Chandola2009AcmComputSurv,Peel2014Aaai,Akoglu2015DataMinKnowlDisc,Masuda2016book,Wang2017ProcIJCAI,Zambon2018IEEETransNNLS}.
A main difference between our system state dynamics and these methods is that detection of system state dynamics is concerned with not only the change, but what is before and after the change. System states are furthermore recurrent by nature. If a state is visited only once in the entire observation period, it would be practically unclear whether to regard it as a state characterising the system or as an anomaly. While change-point detection focus on the change, our approach is more akin to dimensional reduction~\cite{laurence} of the network.

\section{Methods}

\begin{figure}
\includegraphics[width=16cm]{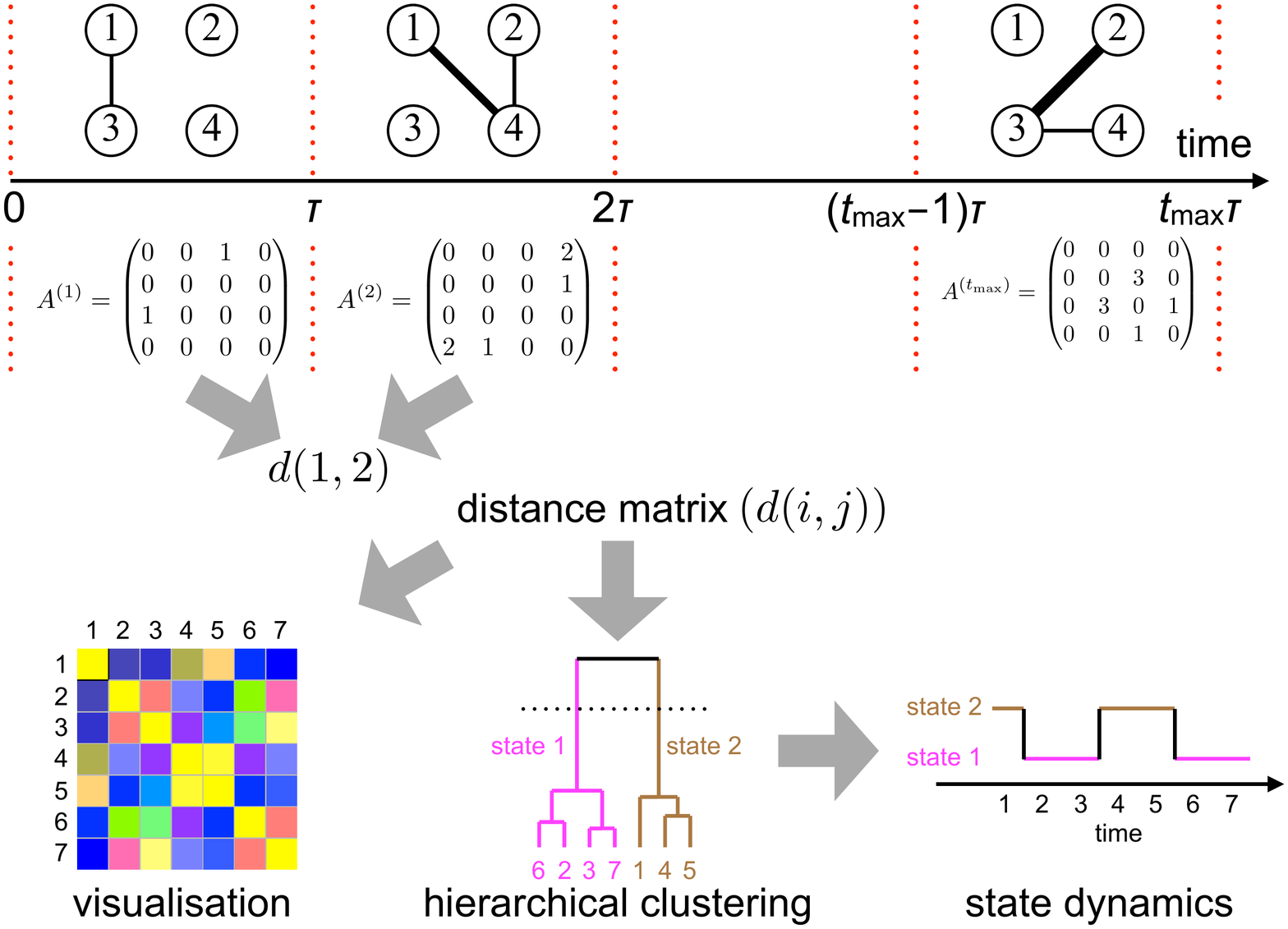}
\caption{Workflow for calculating the system state dynamics of temporal networks. One first specifies the size of the time window, $\tau$, to divide the given temporal network data into a sequence of snapshot networks, or the corresponding adjacency matrices of length $t_{\max}$. Each snapshot network is allowed to be directed and/or weighted. Then, the distance between each pair of snapshot networks, $d(i,j)$ ($1\le i, j\le t_{\max}$), is calculated. Visualisation of the distance matrix or the similarity matrix (which is straightforward to obtain from the distance matrix) would present a recognisable signature of system state dynamics if the temporal network data are inherited with clear system state dynamics. Formally, one runs a hierarchical (or other) clustering method on the distance matrix and regards a cluster as a state. The number associated with the leaf in the dendrogram represents the discretised time. Finally, the system state dynamics are obtained as a time sequence of the states.}
\label{fig:schem}
\end{figure}

The flow of the proposed analysis is schematically shown in Fig.~\ref{fig:schem}.
MATLAB code to calculate and visualise state dynamics for temporal network input data is provided on Github \cite{state-dyn-code}.
First, we transform a given data set of temporal networks into a sequence of static networks, i.e., snapshots. Each snapshot accounts for a non-overlapping time window of length $\tau$ \cite{Masuda2016book}. In other words, the $t$th snapshot network is given by all the events between node pairs that have occurred in time window $[(t-1)\tau, t\tau)$. Second, we measure the pairwise distance between all pairs of snapshot networks. Third, we run a clustering algorithm on the distance matrix to categorise the snapshots into discrete states. Each of these three subroutines, i.e., division of temporal network data into snapshots, a graph distance measure and a clustering algorithm applied to the distance matrix, is conventional. The value of the present method is to combine these pieces together as a single data analysis pipeline to extract state dynamics of temporal networks. We explain the distance measures between static networks, our clustering method and the empirical data sets of temporal networks in the following subsections.

\subsection{Distance measures\label{sub:graph distances}}

To categorise snapshot networks into discrete groups, we need a distance measure for static networks. There are a variety of graph distance measures
\cite{Pincombe2005AsorBull,Livi2013PatAnalAppl,Dedomenico2016PhysRevX,Koutra2016AcmTransKnowlDiscData}. 
We use seven state-of-the-art distance measures explained below. We assume undirected networks, whereas the extension of the proposed methods to the case of directed temporal networks is straightforward.

The \emph{graph edit distance} is
defined as the number of edits needed to transform one network to another
\cite{Pincombe2005AsorBull,Gao2010PatAnalAppl,Livi2013PatAnalAppl}.
Denote the number of nodes and edges in network $G$ by $N(G)$ and $M(G)$, respectively. The graph edit distance between two networks $G_1$ and $G_2$ is given by
\begin{equation}
d = N(G_1) + N(G_2) - 2 N(G_1 \cap G_2) + M(G_1) + M(G_2) - 2 M(G_1 \cap G_2),
\label{eq:graph edit distance}
\end{equation}
where $d$ denotes the distance, $G_1 \cap G_2$ is the network composed of the nodes and edges that commonly exist in $G_1$ and $G_2$. When analysing temporal network data, one usually knows the label of each node that is commonly used across different time points. Therefore, one does not need to match $G_1$ and $G_2$ by permuting nodes. In contrast, the graph edit distance usually involves graph matching. Despite this discrepancy, we decided to use the graph edit distance simply as one of the several graph distance measures to be compared.

DELTACON is a scalable measure of network distance
\cite{Koutra2016AcmTransKnowlDiscData}. It first calculates affinity between all pairs of nodes separately for $G_1$ and $G_2$ using a belief propagation algorithm. Then, one calculates a distance between the affinity matrix for $G_1$ and that for $G_2$. Because our networks are not large in terms of the number of nodes, we use the original version of DELTACON, not a faster approximate version \cite{Koutra2016AcmTransKnowlDiscData}. We use the MATLAB code offered by the authors of Ref.~\cite{Koutra2016AcmTransKnowlDiscData} for computing DELTACON.

The other five distance measures are spectral in nature. In general, spectral distances are based on the comparison between the eigenvalues of the two matrices defining the networks, such as the adjacency or Laplacian matrices \cite{Wilson2008PatRecog}. Because the Laplacian matrices have been shown to be superior to the adjacency matrix when applied to the spectral distance \cite{Wilson2008PatRecog}, we investigate the following five variants of Laplacian spectral distance measures.
First, we distinguish between the combinatorial Laplacian matrix and the normalized Laplacian matrix
\cite{Chung1997book-spectral,Cvetkovic2010book,Masuda2017PhysRep}.
The combinatorial Laplacian matrix is defined by
$L=D-A$, where $A$ is the adjacency matrix and $D$ is the diagonal matrix whose $i$th diagonal element is equal to the degree of node $i$.
The normalised Laplacian matrix is given by
$L^{\prime} = D^{-1/2}LD^{-1/2} = I - D^{-1/2} A D^{-1/2}$, where $I$ is the identity matrix.
We denote the eigenvalues of either type of Laplacian matrix by
$\lambda_1=0 \le \lambda_2 \le \cdots \le \lambda_N$, where $N$ is the number of nodes.

The quantum spectral Jensen-Shannon divergence was proposed as a similarity measure for networks
\cite{Dedomenico2016PhysRevX}. It is an entropy-based measure. To define the entropy, we first define the density matrix by
$\bm\rho = e^{-\beta L} / \sum_{i=1}^N e^{-\beta\lambda_i}$, where $\beta$ is a parameter representing the amount of time for which a diffusion process is run on the network. A large value of $\beta$ implies that the distance between networks is defined based on their differences in relatively global structure. Note that the eigenvalues of $\rho$ sum up to one such that $\rho$ is qualified as a density matrix in the sense of quantum mechanics. The von Neumann entropy is defined by 
$S(\bm\rho) = - \sum_{i=1}^N \tilde{\lambda}_i \log_2 \tilde{\lambda}_i$, where $\tilde{\lambda}_i$ is the $i$th eigenvalue of $\bm\rho$. Given the two density matrices $\bm\rho_1$ and $\bm\rho_2$ that correspond to networks $G_1$ and $G_2$, respectively, the distance measure based on the Jensen-Shannon divergence is given by
\begin{equation}
d = \sqrt{S\left(\frac{\bm\rho_1+\bm\rho_2}{2}\right) - \frac{1}{2}\left[S(\bm\rho_1) + S(\bm\rho_2) \right]}.
\end{equation}

The other four Laplacian spectral distance measures are defined as follows.
For each of $L$ and $L^{\prime}$, we consider the following two types of spectral distance.
The unnormalized spectrum distance for either type of Laplacian matrix is defined by \cite{QiuHancock2006PatRecog,Wilson2008PatRecog}
\begin{equation}
d = \sqrt{\sum_{i=1}^{n_{\rm eig}} \left[\lambda_{N+1-i}(G_1) - \lambda_{N+1-i}(G_2)\right]^2}.
\end{equation}
Remember that $\lambda_{N+1-i}(G)$ is the $i$th largest eigenvalue of the Laplacian matrix of network $G$. The integer $n_{\rm eig}$ is the number of eigenvalues to be considered in the calculation of the distance. We set $n_{\rm eig}=N$.
The normalised spectrum distance for either type of Laplacian matrix is defined by
\begin{equation}
d = \sqrt{\frac{\sum_{i=1}^{n_{\rm eig}} \left[\lambda_{N+1-i}(G_1) - \lambda_{N+1-i}(G_2)\right]^2}
{\max\{ \sum_{i=1}^{n_{\rm eig}} \left[\lambda_{N+1-i}(G_1)\right]^2, \sum_{i=1}^{n_{\rm eig}}
\left[\lambda_{N+1-i}(G_2)\right]^2\}}}.
\label{eq:normalized d}
\end{equation}
Note that the normalisation factor of the denominator on the RHS of Eq.~\eqref{eq:normalized d} is often
given as the minimum rather than the maximum \cite{Pincombe2005AsorBull,Bunke2007book}.
We decided to use the maximum because the use of the
maximum bounds $d$ by $\sqrt{2}$ from above, whereas the distance defined with the minimum can yield arbitrarily large distances.

\subsection{States of temporal networks}

We view a temporal network as a sequence of $t_{\max}$ static networks (i.e., snapshots).
We assign to each of the $t_{\max}$ snapshots a state as follows.
On the distance matrix between the snapshots, $d(i,j)$, where $1\le i,j\le t_{\max}$, we apply a standard hierarchical clustering algorithm and regard each cluster as a state. We used the shortest distance to define the distance between clusters. We used ``linkage'' and ``cluster'' in-built functions in MATLAB with the default option.

The hierarchical clustering provides partitions of the snapshots into states with all possible numbers of states, $C$, i.e., $1\le C\le t_{\max}$. Given that there are many criteria with which to determine the number of clusters \cite{Arbelaitz2013PatRecog}, we determined the final number of states using the Dunn's index defined by \cite{Dunn1974JCybern} 
\begin{equation}
\frac{\min_{1\le c \neq c^{\prime}\le C} \; \min_{i\in c \textrm{th state},\; j\in c^{\prime} \textrm{th state}} d(i,j)}
{\max_{1\le c^{\prime\prime}\le C}\; \max_{i^{\prime},j^{\prime}\in c^{\prime\prime}\textrm{th state}} d(i^{\prime},j^{\prime})}.
\label{eq:Dunn}
\end{equation}
The numerator in Eq.~\eqref{eq:Dunn} represents the smallest distance between two states among all pairs of states. The denominator represents the largest diameter of the state among all states.
We adopt the value of $C$ ($2\le C\le t_{\max}$) that maximises the Dunn's index. Note that some other popular indices for determining the number of clusters in hierarchical clustering, such as the Calinski-Harabasz index \cite{Calinski1974CommStat}, are not applicable because they require the centroid of the data points in a cluster, which is not a priori defined  for networks.

\subsection{Empirical data}

We will use four data sets of empirical temporal networks. All the data sets represent either interaction between a group of people or their physical proximity. We also investigated other similar data sets, such as those used in Ref.~\cite{Holme2017arxiv}, but did not find notable state dynamics. Basic properties of the data sets are shown in Table~\ref{tab:data}.

The primary school data set was gathered by the Sociopatterns organisation using radio-frequency identification devices (RFID). This technology records proximity between humans within about 1.5 m distance. The participants of the study were $232$ primary school children and $10$ teachers in France~\cite{school}. They were recorded over two consecutive days in October 2009, between 8:45am and 5:20pm on the first day and between 8:30am and 5:05pm on the second day. There were $N=242$ nodes.



We used a data set gathered with iMote sensors carried by groups of users in Cambridge, UK. The data set was downloaded from CRAWDAD \cite{crawdad}. We refer to the data set as the Cambridge data set.
%
%
The Cambridge data set was recorded from students belonging to the Systems Research Group in the Computer Laboratory of the University of Cambridge. The recording covered approximately five days between the 25th of January, 2005, Tuesday at 2pm and the 31st of January, 2005, Monday in the afternoon. Twelve participants yielded data without technical problems in their devices. The other nodes correspond to external devices. There are $N=223$ nodes in total.

The Reality Mining data set comes from an experiment conducted on students of Massachusetts Institute of Technology. The students were given smartphones, and their pairwise proximity was recorded via the Bluetooth channel~\cite{Eagle2006PersUbiquitComput}. We use a subset of this data set, which was also used in Ref.~\cite{Scholtes2014NatComm}. The sample had $N=64$ individuals recorded for approximately 8.5 hours.

The Copenhagen Bluetooth data set is a subset of the data described in Ref.~\cite{Stopczynski}. Like the previous data sets, this data set was also recorded via the Bluetooth channel of smartphones. The data were acquired in an over year long experiment where around 1,000 university students were equipped with smartphones, reporting data of their communication over different channels. We use a threshold on the received signal strength indicator of $-75$.  The subset of this data set that we use covers four weeks involving $N=703$ people. For privacy reasons, the exact date and time of the contacts were unavailable to us.


\section{Results}

\subsection{Comparison between different distance measures on classification of static networks} 

Before analysing temporal networks, we first compared static networks generated by different models. The purpose of this analysis was to compare performances of the different graph distance measures introduced in section~\ref{sub:graph distances} in distinguishing static network samples generated by two different models (or one model with two different parameter values). This exercise is relevant to our method because, in our method, snapshots of a temporal network are clustered into groups (i.e., states) according to a measure of graph distance between pairs of snapshots, which are static networks. Therefore, a graph distance measure that yields a good discrimination performance for the static networks in this section is expected to be also a good performer in the next section, where we calculate state dynamics of temporal networks for empirical data sets. We seek such good performers in this section.

In fact, the event rate averaged over all pairs of nodes and a time window is one of the simplest measurements of temporal networks. We are not interested in the states of temporal networks that simply correspond to different event rates. Therefore, in this section, we compared two ensembles of static networks with approximately the same edge density.

First, we compared the regular random graph (RRG) and the Barab\'{a}si-Albert (BA) model \cite{Barabasi1999Science}. We generated RRGs using the configuration model with $N=100$ nodes and the degree of each node equal to six. We generated networks from the BA model with parameters $m_0=3$ and $m=3$, where $m_0$ is the number of nodes that initially form a clique in the process of growing a network, and $m$ is the number of edges that each new node added to the network initially has. These parameter values yield mean degree $\approx 6$, while the degree obeys a power-law distribution with power-law exponent three. In the BA model, the node index is randomly assigned to the $N$ nodes, and therefore the node index is not correlated with the age or the degree of the node. For each distance measure, we carried out the following three comparisons. First, we calculated the distance between two networks generated by the RRG. Second, we calculated the distance between two degree-heterogeneous networks generated by the BA model. Third, we calculated the distance between a network generated by the RRG and a network generated by the BA model. In each case, we generated $10^3$ pairs of networks and calculated the average and standard deviation of the distance between a pair of networks. All the networks used in these comparisons are independent samples. 

\begin{figure}
\includegraphics[width=16cm]{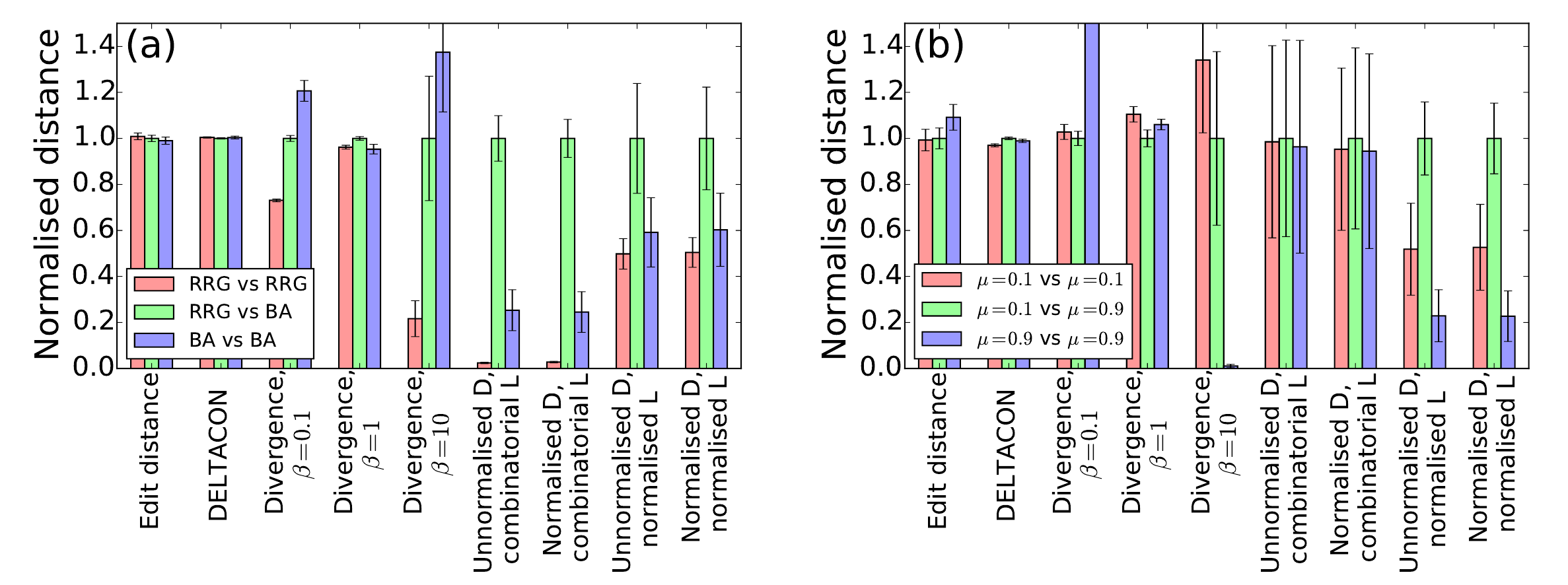}
\caption{Distance between pairs of static networks with seven graph distance measures.
(a) Regular random graphs (denoted by RRG) and the BA model. (b) LFR model with different parameter values. 
The bar graphs present the average of the distance values over $10^3$ pairs of networks sampled from each model. The distance values are normalised by that for the average of the distance between (a) the RRG and BA pair or (b) the LFR network with $\mu=0.1$ and that with $\mu=0.9$. The error bars represent a standard deviation calculated from the $10^3$ pairs of networks.}
\label{fig:between models}
\end{figure}

The values of the distance are shown in Fig.~\ref{fig:between models}(a). The distance values were scaled such that the average distance between the RRG and BA model corresponds to one in the figure. 
The figure indicates that, for the edit distance and DELTACON, the distance between the RRG and BA model is not significantly larger than  between a pair of RRGs and between a pair of BA networks. These distance measures would take small values when edges between particular pairs of nodes exist in both of the two networks. Because the RRG and BA models are stochastic algorithms, this situation does not often happen even for a pair of networks generated by the same model. In applications to temporal networks, it is probably too demanding to require that events between the same pair of nodes should happen in many node pairs for two snapshots to be judged to be similar. The Jensen-Shannon divergence discriminates the RRG and BA when $\beta=1$ but not when $\beta=0.1$ or $\beta=10$. The other four spectral distance measures discriminate the RRG and the BA model in the sense that the RRG-BA distance is significantly larger than the RRG-RRG and BA-BA distances.

We next turn to another pair of models, which are the Lancichinetti-Fortunato-Radicchi (LFR) benchmark with different parameter values. The LFR benchmark generates networks with community structure \cite{Lancichinetti2009PhysRevE-benchmark}.
The model creates networks having a heterogeneous degree distribution and a heterogeneous distribution of community size. 
Parameter $\mu$ tunes the extent of mixing of different communities, such that a fraction $\mu$ of edges incident to each node goes to different communities. A small value of $\mu$ implies a strong community structure.
We compare networks generated with $\mu=0.1$ and those generated with $\mu=0.9$. We set $N=100$, the mean degree to six, the largest degree to $N/4 = 25$, the power-law exponent for the degree distribution to two and the power-law exponent for the distribution of community size to one. It should be noted that the degree distribution is independent of $\mu$. Therefore, the two groups of networks to be compared differ in the strength of the community structure but not in the degree distribution, presenting a more difficult classification problem than the previous one. For example, the edges of a social temporal network in a primary school \cite{school} are mostly confined within classes during the classroom time, corresponding to strong community structure. In contrast, pupils tend to be mixed across classes during playtime, probably corresponding to weak community structure.

The network distance is compared between pairs of networks generated with $\mu=0.1$ and $0.9$ in Fig.~\ref{fig:between models}(b).
The discrimination between $\mu=0.1$ and $\mu=0.9$
is unsuccessful with the edit distance in the sense that the distance between $\mu=0.1$ and $\mu=0.9$ is not statistically larger than that between a pair of LFR networks generated with the same $\mu$ values. With DELTACON, the distance between $\mu=0.1$ and $\mu=0.9$ is significantly larger than the other two cases. However, the values of the distance are close among the three cases.
The Jensen-Shannon spectral divergence also fails for the three values of $\beta$. The unnormalised and normalised spectral distances are not successful when they are combined with the combinatorial Laplacian matrix, which is presumably for the following reason. The community structure and the lack thereof are reflected by the small eigenvalues of the Laplacian matrix \cite{Newman2004EurPhysJB,Arenas2006PhysRevLett}. In contrast, the largest eigenvalue of the combinatorial Laplacian matrix is proportional to the largest node degree in the network \cite{Fiedler1973CzechMathJ}. Because the largest degree depends on samples of networks, the fluctuation in the largest eigenvalue would be a dominant contributor to the spectral distance. Then, the spectral distance under-represents the discrepancy between small eigenvalues in the two networks, which are related to community structure, in particular for degree-heterogeneous networks. In contrast, Fig.~\ref{fig:between models}(b) indicates that the spectral distances applied to the normalised Laplacian matrices discriminate between LFR networks with $\mu=0.1$ and those with $\mu=0.9$.
It should be noted that the largest eigenvalue of the 
normalised Laplacian matrix does not scale with the largest degree of the node in the network, i.e.,
$\lambda_N\le 2$ \cite{Chung1997book-spectral,Cvetkovic2010book}.

Given the results shown in Fig.~\ref{fig:between models}, in the next section, we will focus on the two spectral distance measures applied to normalized Laplacian matrices of snapshots of temporal networks.


\subsection{Extracting states in temporal networks}

In this section, we examine state dynamics in empirical temporal networks. For each data set, we have to specify the duration of the time window, $\tau$, to partition the temporal network into a sequence of $t_{\max}$ snapshots. The choice of $\tau$ is arbitrary and is shown in Table~\ref{tab:data}.
We calculate the unnormalised or normalised spectrum distances between each pair of normalised Laplacian matrices corresponding to two snapshots. For visualisation purposes, we transform the distance matrix to a similarity matrix, where the similarity between two snapshots $t_1$ and $t_2$ is defined as
$\text{sim}(t_1, t_2) = 1 - d(t_1, t_2) /\max_{1\le t^{\prime}, t^{\prime\prime}\le t_{\max}} d(t^{\prime},t^{\prime\prime})$, such that $0\le \text{sim}(t_1, t_2)\le 1$.
The distance value of zero between two snapshots corresponds to the similarity value of one. 

\begin{figure}
\includegraphics[width=16cm]{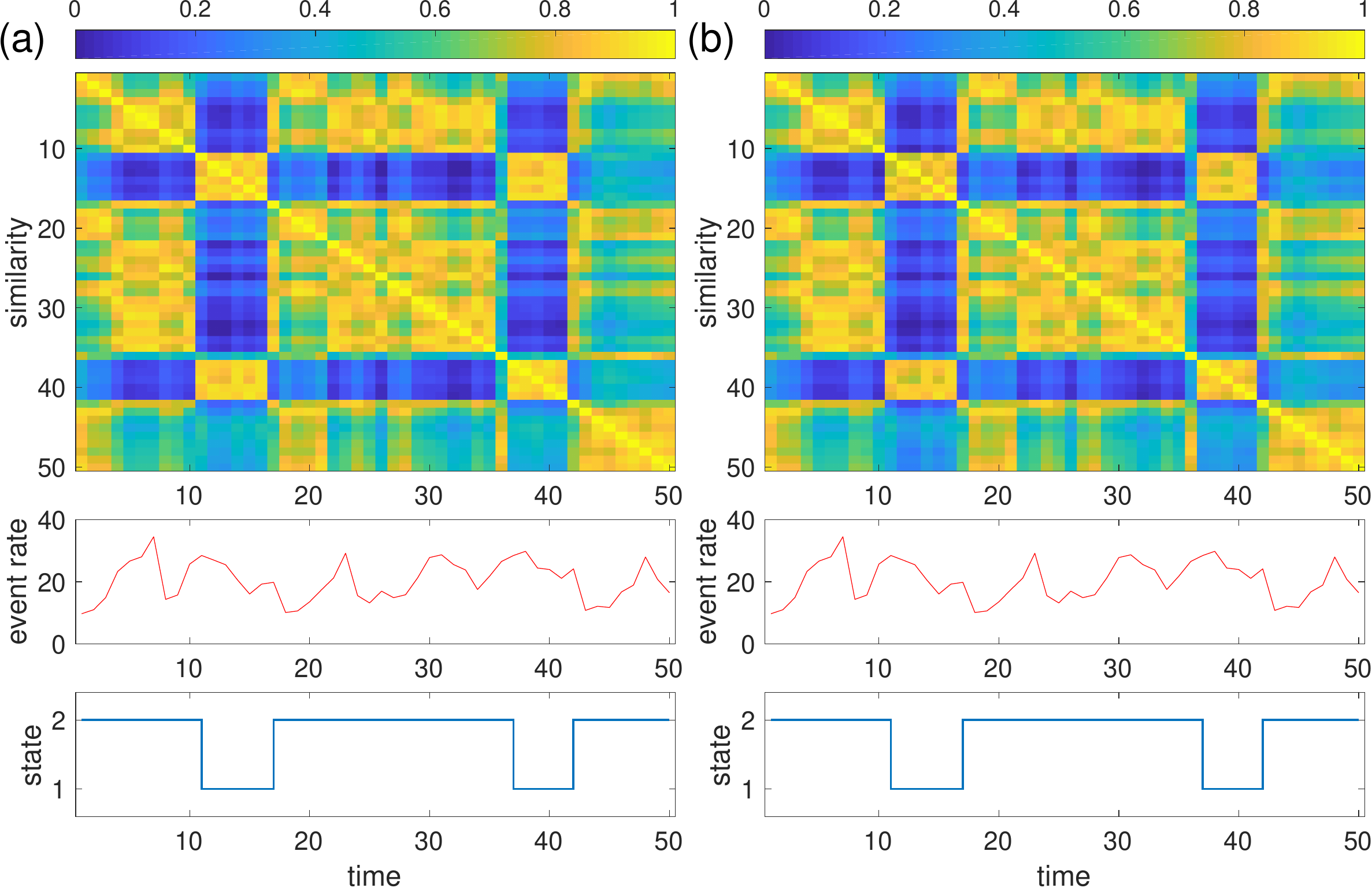}
\caption{System state dynamics for the primary school data. (a) Unnormalised spectral distance. (b) Normalised distance. The top panels show similarity matrices. The middle panels show the number of events per individual in each time window. The bottom panels show the system state dynamics. The index is the discrete time, each one corresponding to a time window. The length of a time window is $\tau = 20$ min.}
\label{fig:primary}
\end{figure}

The results for the primary school data for the unnormalised and normalised spectral distances are shown in Fig.~\ref{fig:primary}(a) and Fig.~\ref{fig:primary}(b), respectively. A snapshot accounts for $\tau = 20$ min. Each of the two consecutive days spans 25 time windows. The results for the two days are concatenated in the figure.
The upper matrix  is the similarity matrix. We notice that 
the snapshots during the lunch time are close to each other and across the two days. The snapshots in other periods are also close to each other, albeit to a lesser extent. The average number of events for an individual in a time window is shown in the middle panel. The lunch-time snapshots are not characterised by event rates that are different from those in different time windows. The panel to the bottom is the time series of the state of the temporal network. Dunn's index suggested two states, i.e., a lunch-time state and a class-time state, with both spectral distance measures. The present results are consistent with an analysis of a single day of the same data set using graph signal processing \cite{Hamon2016IeeeTransSignalInfoProcNetw}. That study showed that different modes are dominant in the lunchtime and in the other times of the day.

\begin{figure}
\includegraphics[width=16cm]{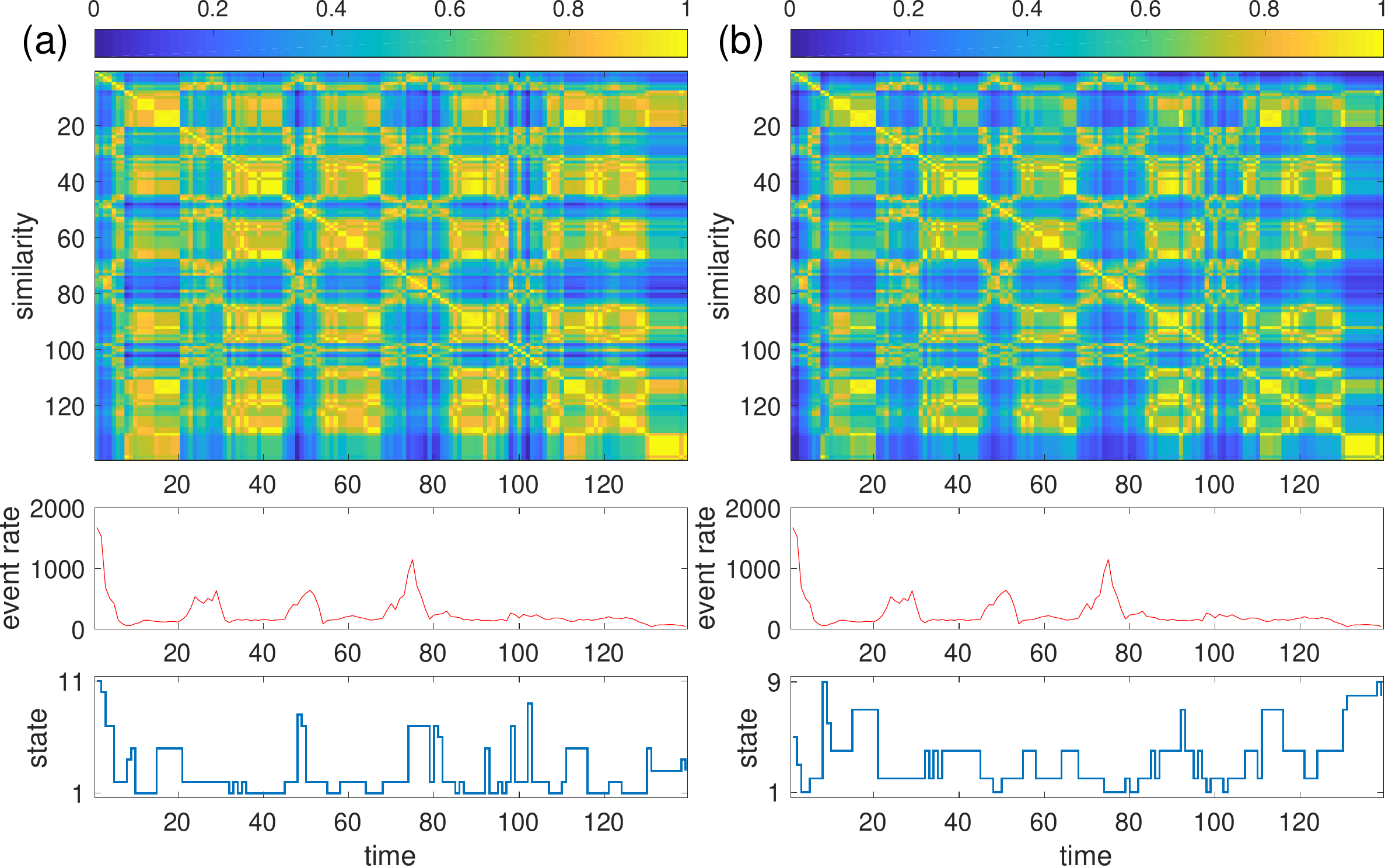}
\caption{Results for the Cambridge data. (a) Unnormalised spectral distance. (b) Normalised spectral distance.
We set $\tau = 1$ hour.}
\label{fig:cambridge}
\end{figure}

Similar results were obtained for the Cambridge data set with $\tau=1$ hour. We excluded the last six hours because the average event rate was extremely low during the period. Both the unnormalised and normalised spectral distances concluded relatively many states
(Fig.~\ref{fig:cambridge}(a)).
Some of the states are characterised by different mean activity rates (large event rates in states 7, 8, 10 and 11 and small event rates in state 3 and 4 in Fig.~\ref{fig:cambridge}(a); large event rates in states 3 and 5 and small event rates in states 6, 8 and 9 in Fig.~\ref{fig:cambridge}(b)). However, other states are not simply characterised by the mean event rate. For example, state 2 in Fig.~\ref{fig:cambridge}(a), equivalently, state 2 in Fig.~\ref{fig:cambridge}(b), is composed of earlier snapshots with elevated event rates and later snapshots with un-elevated event rates. In addition, state 5 in Fig.~\ref{fig:cambridge}(a), equivalently, state 7 in Fig.~\ref{fig:cambridge}(b), appears in two chunks of time that are separated by many hours. In fact, the snapshots in the two chunks are fairly close to each other in the spectral distance. 

\begin{figure}
\includegraphics[width=16cm]{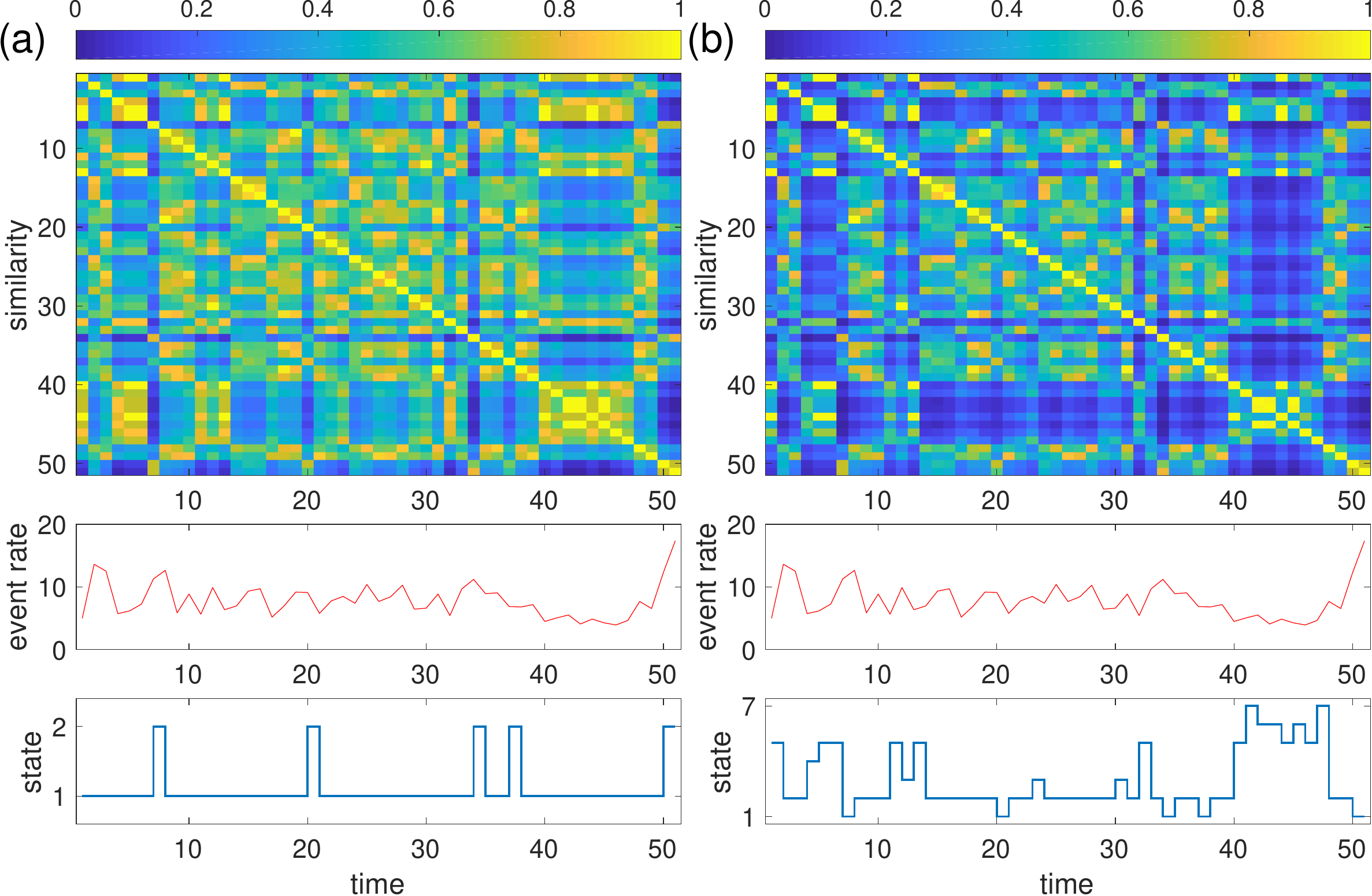}
\caption{Results for the Reality Mining data. (a) Unnormalised spectral distance. (b) Normalised spectral distance. We set $\tau = 10$ min.}
\label{fig:reality}
\end{figure}

The results for the Reality Mining data with $\tau = 10$ min are shown in Fig.~\ref{fig:reality}. With the unnormalised spectral distance, the network occasionally and briefly visits state 2, which is not characterised by a changed mean event rate
(Fig.~\ref{fig:reality}(a)). This result is consistent with a visual inspection of the similarity matrix.
The normalised spectral distance identifies the same state (state 1 in Fig.~\ref{fig:reality}(b)) and also other states. In particular, state 5 is composed of several early snapshots and some snapshots in time windows from 39 to 46, which stand out in the similarity matrix. 

\begin{figure}
\includegraphics[width=16cm]{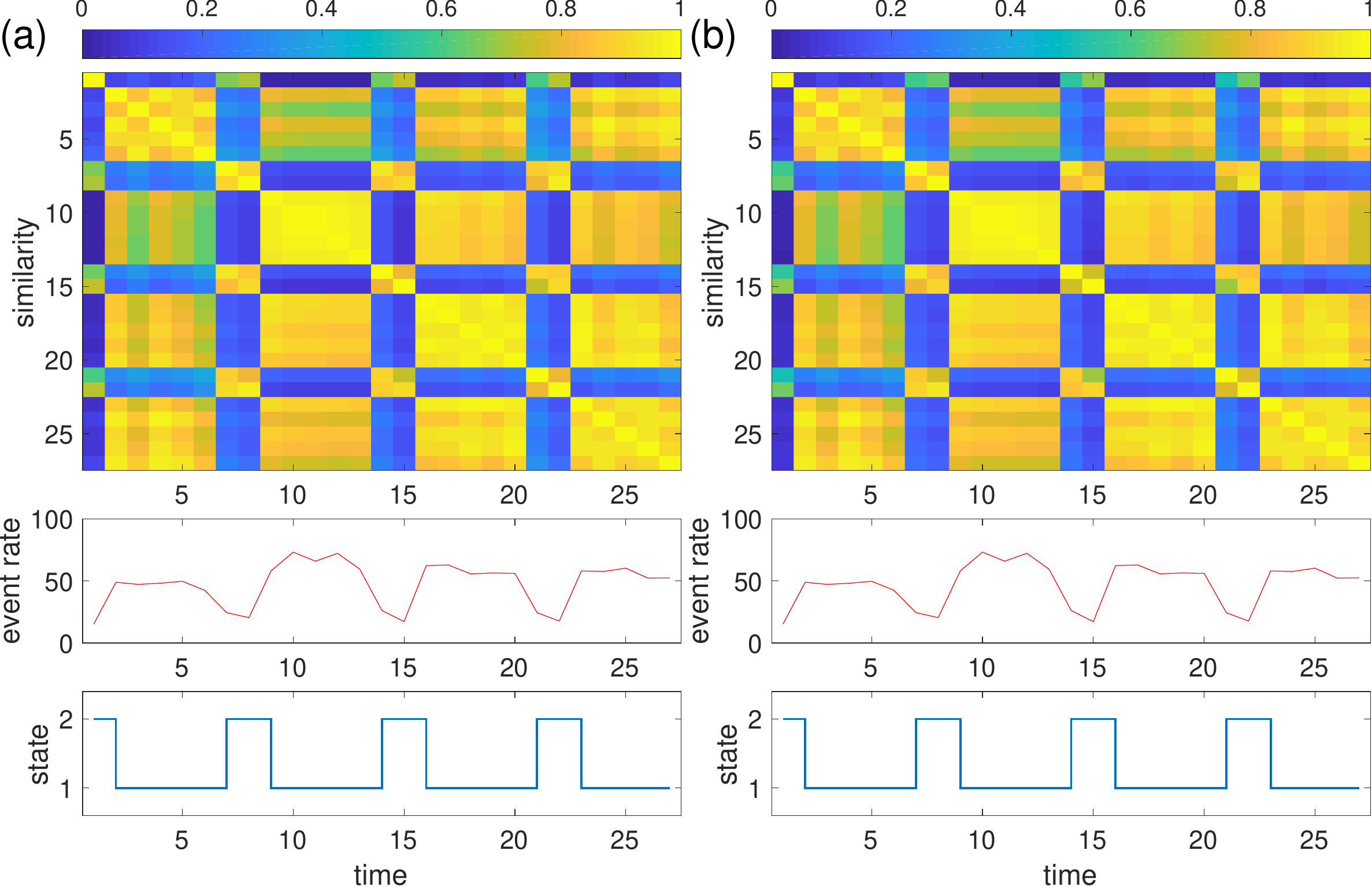}
\caption{Results for the Copenhagen Bluetooth data. (a) Unnormalised spectral distance. (b) Normalised spectral distance. We set $\tau = 1$ day.}
\label{fig:sune-bt75}
\end{figure}

The results for the Copenhagen Bluetooth data with $\tau = 1$ day are shown in Fig.~\ref{fig:sune-bt75}. For both spectral distance measures, the similarity matrix suggests that the weekdays and weekends constitute distinct states. The hierarchical clustering identifies these two states with either distance measure. It should be noted that the average event rate provides sufficient information for one to distinguish between the weekdays and the weekends; the communication is considerably sparser on weekends than weekdays.


\section{Discussion}

We proposed a methodology to use graph similarity scores to construct a sequence of states of temporal networks. We tested this framework with two spectrum distances combined with the normalized Laplacian matrix. Across different data sets, the method revealed states of the temporal networks.
Some networks were categorised into discrete states although the event rate was not specifically modulated over time. 

\begin{figure}
\includegraphics[width=16cm]{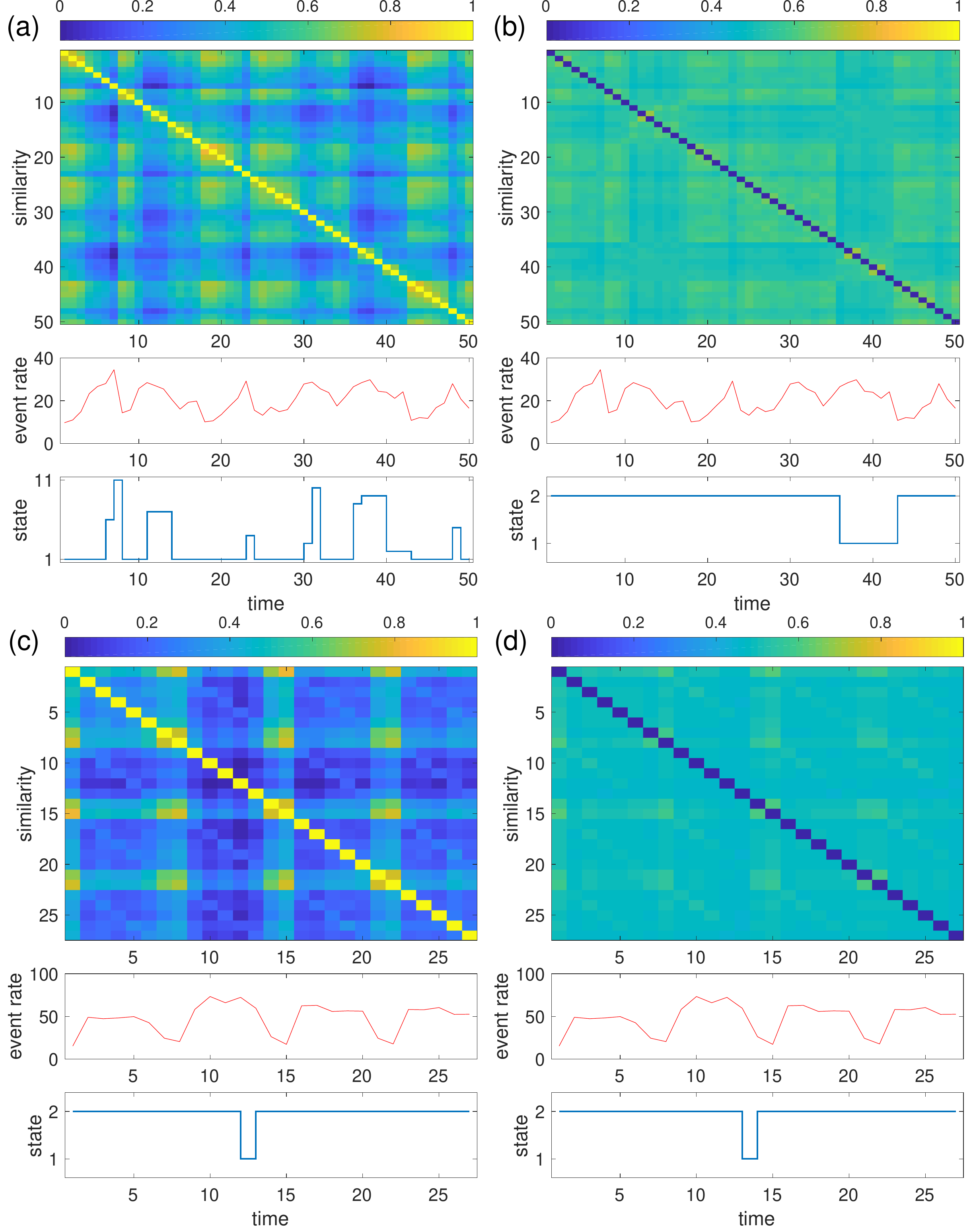}
\caption{System state dynamics with the graph edit distance and DELTACON. (a) Primary school data with the edit distance. (b) Primary school data with DELTACON. (c) Copenhagen Bluetooth data with the edit distance. (d) Copenhagen Bluetooth data with DELTACON. 
The length of a time window is $\tau = 20$ min and $\tau=1$ day for the primary school data and Copenhagen bluetooth data, respectively. We observe that the system state dynamics are much more blurred with the edit distance or DELTACON than with the Laplacian spectral distances (Figs.~\ref{fig:primary} and \ref{fig:sune-bt75}).}
\label{fig:SI editdist and deltacon}
\end{figure}

The present study has not systematically investigated distance measures for networks.
A consistent finding in the present study is that distance measures based on the comparison of individual nodes and edges (i.e., graph edit distance and DELTACON) are probably too stringent. It was the case in the comparison between static network models (Fig.~\ref{fig:between models}). We additionally confirmed that the edit distance and DELTACON performed poorly for the primary school and Copenhagen Bluetooth data, in which the states were relatively clear-cut and interpretable (Fig.~\ref{fig:SI editdist and deltacon}). Apart from that, the present method can be combined with other graph distance measures such as graph kernels \cite{Livi2013PatAnalAppl}, graph embedding \cite{Livi2013PatAnalAppl} or those based on feature vectors \cite{Tsuda2006ProcICML,Berlingerio2013ProcAsonam}. Our spectral distances implicitly ignore the node identities when comparing snapshots. In practice, this information is often available. In such cases, graph distance measures that take the node identity into account may yield better results.  For example, state dynamics may be induced by activation of different network communities at different times. If so, a graph distance measure that exploits community structure of networks \cite{Onnela2012PhysRevE} may yield better results than with the Laplacian or other graph distances.

Many systems are composed of different layers of networks
\cite{Boccaletti2014PhysRep,Kivela2014JCompNetw}. There are algorithms to categorise individual layers of a multilayer network into groups \cite{Dedomenico2016PhysRevX,Iacovacci2016Chaos}. 
A temporal network can be regarded as a multilayer network if one regards each snapshot network as a layer.
Therefore, these previous methods are directly applicable to the current framework. 
On the other hand, a more standard approach to regard a temporal network as a multilayer network is
to connect two snapshots of the temporal network if and only if they are consecutive in time \cite{Mucha2010Science,Boccaletti2014PhysRep,Kivela2014JCompNetw,Masuda2016book}.
In the present study, we did not connect snapshots across different times. Doing so would bind temporally close snapshots  into the same state such that the system state dynamics would experience less switching. Introducing inertia to the state dynamics by connecting consecutive snapshots or by other means may be useful to enhance interpretability of the results for some data.

The idea of system state dynamics in temporal networks has recently been advocated for time-varying neuroimaging data, called chronnectome \cite{Calhoun2014Neuron,ChoeNebel2017Neuroimage}. For those data, networks are correlational, so-called the functional networks, and are composed of brain regions of interest used as nodes and the correlation value (or its thresholded version) conventionally used as edges \cite{Bullmore2009NatRevNeurosci,Sporns2011book,Bassett2017NatNeurosci}.
Chronnectome analysis has revealed, for example, different patterns in system state dynamics between patients and controls
\cite{Calhoun2014Neuron,Rosch2018NetwNeurosci}. The present framework can be regarded as chronnectome for general temporal networks including non-correlational ones, with general graph distance measures. Its applicability is not limited to social or neuronal temporal networks.
For example, protein-protein interaction networks are also suggested to be dynamic, where ``date-hub'' proteins choreograph temporality of networks by binding different partners at different times and locations \cite{Han2004Nature,ChangXu2013SciRep}. Ecological interspecific interaction networks are also dynamic, and the network dynamics affect stability of an ecosystem \cite{Ushio2018Nature}. Our method applied to these and other systems may tell us the state of the system at each time point as well as the function of the system associated with the individual states.

A small number of leading principal components of time series obtained from human behaviour can predict much of the behaviour of the individual. Such principal components are termed the eigenbehaviours \cite{Eagle2009BehavEcolSociobiol}. This method is orthogonal to our approach. In the eigenbehaviour analysis, each eigenbehaviour, i.e., principal component, is the time series of behaviour, and hence is derived from the entire observation time window. Therefore, an eigenbehaviour, if measured for a temporal network, will be spread over time in general. In particular, different eigenmodes may be simultaneously active. In contrast, our method is a partition of the time axis into discrete states. 

The present algorithm and its variants can be applied to adaptive networks, in which, by definition, nodes change their behaviour in response to the system state of the network and dynamic processes occurring on it (e.g., epidemic processes) \cite{GrossSayama2009book,Sayama2013ComputMathAppl}.
If an adaptive network and the dynamical process on top of it evolve towards an equilibrium, state dynamics are irrelevant except in the transient because the system state will not change forever after the transient. However, if the eventual dynamics are of non-equilibrium nature, the present method may be able to find transitions between distinct states that characterise the network dynamics in a dynamic equilibrium. One example is the ``diplomat's dilemma'', in which agents simultaneously try to achieve a high centrality value and low degree. Ref.~\cite{Holme2006PhysRevLett} shows how these conflicting optimisation criteria (because the degree and centrality are usually positively correlated) lead to a situation where the system can undergo sudden structural reorganisations after long periods of quiescence. In this model, the present method may detect active and quiescent periods as distinct states.

A limitation of the present approach is the assumption that the entire system can be described by a single system state. In fact, we found nontrivial interpretable results in only one of the four data sets that we investigated (i.e., primary school data set). In many cases, one could argue that it makes more sense to describe different groups in the data as having their own system state. If our method is complemented by a community detection component, this problem could be circumvented. However, community detection in temporal networks is notoriously hard to tackle with a principled approach~\cite{rossetti}.
Another limitation is arbitrariness in the choice of the subroutines and parameter values. For example, there are myriads of distance measures for networks \cite{Livi2013PatAnalAppl,Koutra2016AcmTransKnowlDiscData}. In addition, hierarchical clustering comes with various options in how to connect different clusters, and there are many other data clustering methods \cite{Duda2001book}. Furthermore, there are other methods for determining the number of groups apart from the Dunn's index \cite{Arbelaitz2013PatRecog}. Our choice of the length of the time window was also arbitrary. The purpose of the present study was to present a new characterisation of temporal networks. More exhaustive examinations of these variations  are left as future work.

\section*{Acknowledgments}

We thank Sune Lehmann for discussion and providing the Copenhagen Bluetooth data set. We thank Lorenzo Livi for discussion.
N.M. acknowledges the support provided through JST CREST Grant Number JPMJCR1304.
P.H. was supported by JSPS KAKENHI Grant Number JP 18H01655.

\section*{Author contributions}

N.M. conceived the research; N.M. and P.H. designed the research; N.M. performed the computational experiments; N.M. and P.H. wrote the paper.

\section*{Additional information}

\subsection*{Competing interests}

The authors declare no competing interests.


\newpage
\clearpage

\newpage
\clearpage

\newpage
\clearpage

\newpage
\clearpage





\newpage
\clearpage

\newpage
\clearpage

\newpage
\clearpage

\newpage
\clearpage

\begin{table}
\centering
\caption{Properties of the temporal network data. Variable $N$ represents the number of nodes; ``Resolution'' indicates the time resolution with which the data were originally collected; $\tau$ is the length of the time window to create snapshot networks; $t_{\max}$ is the number of time windows.}
\label{tab:data}
\bigskip
\begin{tabular}{ccccc}\hline
Data & $N$ & Resolution & $\tau$ & $t_{\max}$ \\ \hline\hline
Primary school & 242 & 20 s & 20 m & 50 \\
Cambridge & 223 & 1 s & 1 h & 145\\
Reality Mining & 64 & 5 s & 10 m & 51\\
Copenhagen Bluetooth & 703 & 1 s & 1 d & 27\\ \hline
\end{tabular}
\end{table}

\end{document}